\documentstyle[12pt,epsf,epsfig]{article}

\newcommand \be {\begin{equation}}
\newcommand \bea {\begin{eqnarray}}
\newcommand \ee {\end{equation}}
\newcommand \eea {\end{eqnarray}}
\newcommand \eps {\epsilon}

\def \a{\alpha}
\def \b{\beta}
\def \eps{\epsilon}
\def \d{{\delta}}
\def \c{\gamma}

\def \be{\begin{equation}}
\def \ee{\end{equation}}
\def \bea{\begin{eqnarray}}
\def \eea{\end{eqnarray}}
\def \ln{{\rm ln}}

\def \w{\omega}

\def \la{\langle}
\def \ra{\rangle}

\begin{document}
\title{1-loop contribution to the dynamical exponents in spin glasses}
\author{G. Parisi and P. Ranieri}
\maketitle
\begin{center}
Dipartimento di Fisica Universit\`a di Roma La Sapienza and
\vspace{.1in}\\
INFN sezione di Roma I 
Piazzale Aldo Moro, Roma 00185

\begin{abstract}
We evaluate the corrections to the mean field values of the $x$ and $z$ 
exponents at the first order in the $\eps$-expansion, for $T=T_c$. 
We find that both $x$ and $z$ are decreasing when the space dimension 
decreases. 
\end{abstract}
\end{center}
\hyphenation{dia-go-na-li-za-tion}
We want to investigate the purely relaxational dynamics of a short-range 
spin glass model for $T\rightarrow T_c^+$, in the framework of the 
$\eps$-expansion.  The  dynamical properties of the model in the mean field 
theory are very different from those of the models whose dynamical 
properties are usually investigated in the literature. These new feature make 
the computation of the dynamical critical exponents much more involved than 
that of the usual.

In a previous work, \cite{ra.}, we evaluated the Gaussian dynamical 
fluctuations of the order parameter around the MF limit.  The aim of 
this letter is to pursue this analysis, by considering the 1-loop 
correction to the mean field (MF) theory in a renormalization group            
calculation.  At the first order in $\epsilon$, unlike Zippelius 
\cite{z.}, we find results that disagree with the conventional Van 
Hove theory, because we obtain a correction to the kinetic coefficient 
already to the lowest order in the loop expansion.  In this work, 
first we evaluate the correction to MF value of the critical exponents 
$x$, that describes the critical slowing down of the dynamical order 
parameter at the critical point, then we evaluate the correction to 
the $z$ exponent that describes the critical slowing down of the 
dynamical spin glass susceptibility $\chi_{SG}$.  Finally, we check 
that the scaling low, which connect these two exponents, is verified.

We study the soft-spin version of the EA model given by the Hamiltonian:
\be
\b{\cal H}=-\b\sum_{\la ij\ra}J_{ij}s_i
s_{j}+\frac{1}{2}r_o\sum_{i}s_i^2+\frac{1}{4!} g\sum_i s_i^4\; ,
\ee
where $J_{ij}$ are random Gaussian interactions between the nearest-neighbors
sites, with zero average and mean square fluctuations $[(J_{ij})^2]=j^2/n$ 
($n$ is the coordination number).
Purely relaxation dynamics is introduced by the Langevin equation:
\be
\Gamma_0^{-1}\frac{\partial s_{i}(t)}{\partial t}=-\frac{\partial(\b{\cal
H})}{\partial s_i(t)}+\eta_i(t),
\ee
where $\eta$ is the usual Gaussian noise with zero average and variance  
$\la\eta_i(t)\eta_j(t')\ra=\frac{2}{\Gamma_0}\delta_{ij}\delta(t-t')$.
The interesting physical quantities in MF theory are the averaged response 
and correlation functions, defined respectively:
\bea
\label{g1}
\overline{G}(t-t')=\left[\frac{\partial\la s_i(t)\ra_{\eta}}
{\partial h_i(t')}\right]_J&&\\
\label{c1}
\overline{C}(t-t')=\left[\la s_i(t)s_i(t')\ra_{\eta}\right]_J,&&
\eea
where the angular brackets $\la ..\ra_{\eta}$ refer to averages over the noise
and the square brackets $\left[..\right]_J$ over quenched disorder.

Moreover, considering the Gaussian fluctuation, we can define the 
dynamical spin glass susceptibility as follow: 
\be
\label{su1}
\chi_{SG}(i-j;t_3-t_1,t_2-t_4)=\left[\frac{\partial\la s_i(t_3)\ra_\eta}
{\partial h_j(t_1)}
\frac{\partial\la s_i(t_2)\ra_\eta}{\partial h_j(t_4)}\right]_J.
\ee

The dynamical scaling implies that the decay of $C(t)$ is governed by a
characteristic time $\tau$, which diverges at $T_{c}$, as,
for long $t$, we can write:
\be
C(t)=\frac{1}{t^x}\tilde{q}^+(t/\tau),
\ee
where $\tilde{q}^+$ is the universal scaling function in the region 
$T\rightarrow T_c^+$. 
The relaxation time divergence, at the critical point, is connected to 
the correlation length divergence through the dynamical exponent $z$: 
\be
\tau\propto\xi^z.
\ee
According to the scaling hypothesis, 
this exponent is related to the slowing down of the spin glass susceptibility,  
and we expect to have:
 \be
\chi_{SG}(k,\w)=\w^{\frac{2-\eta}{z}}\tilde{f}(k^z/\w)
\ee

The MF behaviour of this model ($n=N$, number of spins, long-range limit),  
in the critical region, is well known \cite{s.z.1}, \cite{s.z.2}, \cite{s.z.3}. 
In the low-frequency limit the response and correlation functions are 
respectively:
\bea
\label{g2} 
\overline{G}(\omega)=\left(1-\sqrt{-i\omega}\right),&& \\
\label{c2}
\overline{C}(\omega)=\frac{2}{\sqrt{i\omega}+\sqrt{-i\omega}},&&  
\eea
while, in the Gaussian approximation, the spin glass susceptibility is
\be
\chi_{SG}(k,\w_1,\w_2)=\frac{1}{k^2+\sqrt{-i\w_1}+\sqrt{-i\w_2}}.
\ee 
The MF value of the  dynamical critical exponents $x$ and $z$, as known, 
are $1/2$ and $4$ respectively.
 
To deal with Langevin disordered dynamic theory as usual we use the dynamic 
functional integral method \cite{dom.pel.}.  
In this formalism, it is conventional 
to introduce an auxiliary field $\hat{s}_i(t)$ and to define an effective 
Lagrangian of an Hubbard-Stratanovich field $Q_i^{\a\b}(t,t')$, \cite{s.z.2}, 
\cite{z.}, such as:
\begin{equation}
\label{q}
2
\displaystyle{\langle Q_{k=0}^{\alpha\beta}(t_1,t_2)
\rangle_{L(Q_{\alpha\beta})}}
=\displaystyle{[\langle\phi_{i}^{\alpha}(t_1)\phi_{i}^{\beta}(t_2)
\rangle_{\eta}]_{J}}
\end{equation}
where the two component vector field is defined:
\be
\phi_{i}^{\a}=(i\hat{s}_i, s_i).
\ee
In considering the correction to MF approximation we derive the following
Lagrangian, as series expansion around the order parameter saddle-point 
value $\overline{Q^{\a\b}}(t,t')$, \cite{s.z.3}, \cite{z.}, \cite{ra.}:    
\bea
&\displaystyle{\hspace{-1.5cm}L(\delta Q^{\a\b})
=-\sum_{t_1,t_2}\sum_{i,j}\tilde{K}^{-1}_{i,j}
\delta Q_{i}^{\alpha\beta}(t_1,t_2)
  A^{\alpha\beta\gamma\delta}\delta Q_{j}^{\gamma\delta}(t_1,t_2)+}&\nonumber\\
&\displaystyle{\hspace{-1cm} \frac{1}{2}\sum_{t_1,t_2,t_3,t_4}\sum_{i}
\delta Q_{i}^{\alpha\beta}(t_1,t_2)
  C^{\alpha\beta\gamma\delta}(t_1,t_2,t_3,t_4)
\delta Q_{i}^{\gamma\delta}(t_3,t_4)+}&\nonumber\\
 &\displaystyle{\hspace{-1cm} \frac{1}{3!}\sum_{t_1,t_2,t_3,t_4,t_5,t_6}\sum_{i}
C^{\alpha\beta\gamma\delta\mu\nu}(t_1,t_2,t_3,t_4,t_5,t_6)
  \delta Q_{i}^{\alpha\beta}(t_1,t_2)
  \delta Q_{i}^{\gamma\delta}(t_3,t_4)\delta Q_{i}^{\mu\nu}(t_5,t_6)\;.}&
\nonumber\\
\label{zipp}
\eea
For the structure and the meaning of each term of (\ref{zipp}), 
the reader is referred to \cite{ra.}.  
The non local connected propagators of the theory are defined as follow:
\begin{eqnarray}
&\displaystyle{G^{\alpha\beta\gamma\delta}(i-j;t_1,t_2,t_3,t_4)=
[\langle\phi_{i}^{\alpha}(t_1)\phi_{i}^{\beta}(t_2)\phi_{j}^{\gamma}(t_3)
\phi_{j}^{\delta}(t_4)\rangle_{\eta}]_{J}-}&\nonumber\\
&\displaystyle{\hspace{2cm}
[\langle\phi_{i}^{\alpha}(t_1)\phi_{i}^{\beta}(t_2)\rangle_{\eta}]_{J}
[\langle\phi_{j}^{\gamma}(t_3)\phi_{j}^{\delta}(t_4)\rangle_{\eta}]_{J}}& 
\nonumber\\
&\displaystyle{\hspace{-1cm}=4\sum_{l}(\tilde{K}^{-1})_{il}
\sum_{k}(\tilde{K}^{-1})_{jk}
  \langle\delta Q_{l}^{\alpha\beta}(t_1,t_2)
\delta Q_{k}^{\gamma\delta}(t_3,t_4)
\rangle_{L(Q_{\alpha\beta})}}&\nonumber\\
&\displaystyle{\hspace{-1cm}-2(\tilde {K}^{-1})_{ij} 
A^{\alpha\beta\gamma\delta}\delta(1-3)
\delta(2-4)\;.}&
\end{eqnarray}
In \cite{ra.} we evaluated the critical behaviour of these propagators 
for each combination of the indices $\a,\b,\c,\d$ and in any time interval, 
when the cubic interactions is vanishing. 
We write down the general structure of $G^{2211}(\w_1,\w_2,\w_3,\w_4)$
which is present in several of the following 1-loop functions:
\bea
&\displaystyle{
G^{2211}(k;\w_1,\w_2,\w_3,\w_4)=\left[{\d (\w_1+\w_3) \d
(\w_2+\w_4)+\d (\w_2+\w_3) \d (\w_1+\w_4)}\right]}&\nonumber\\
&\displaystyle{\hspace{-.8cm}\tilde{G}^{2211}(k;\w_1\w_2)
+\d (\w_1+\w_2+\w_3+\w_4)
\tilde{\tilde{G}}^{2211}(k;\w_1,\w_2,\w_3,\w_4)}&
\eea
where:
\bea
&\displaystyle{\hspace{-2cm}\tilde{G}^{2211}(k;\w_1,\w_2)=
\frac{1}{ck^2+\sqrt{-i\w_1}+\sqrt{-i\w_2}}}&\\
&\displaystyle{\hspace{-2cm}\tilde{\tilde{G}}^{2211}(k;\w_1,\w_2,\w_3,\w_4)=
\left(\overline{C}(\w_1)\tilde{G}^{2211}(k;-\w_1,\w_2)+
\overline{C}(\w_2)\tilde{G}^{2211}(k;\w_1,-\w_2)\right)\times}&\nonumber
\\&\displaystyle{\hspace{-2cm}\tilde{G}^{2211}(k;\w_1,\w_2)
\tilde{G}^{2211}(k;-\w_3,-\w_4)}g_r(k;\w_1+\w_2)&\\
&\displaystyle{g_r(k;\w_1+\w_2)=\left(ck^2+\sqrt{-i(\w_1+\w_2)}\right)
                   F_1\left(\frac{k^2}{(\w_1+\w_2)^{1/2}}\right)};& 
\eea 
where $F_1(x)$ is an homogenous function of the $\frac{k^2}{\sqrt{\w}}$.
We recall that in this formalism the time-dependent spin glass 
susceptibility (\ref{su1}) is: 
\be
\label{su2}
\displaystyle{\hspace{-3cm}\chi_{SG}(i-j;t_3-t_1,t_2-t_4)=\left[\la
s_i(t_3)\eta_j(t_1)\ra_\eta\la s_j(t_2) \eta_i(t_4)\ra_\eta\right]_j=
\tilde{G}^{1221}(i-j;t_3-t_1,t_2-t_4).}
\ee

Let us consider the 1-loop correction to the "free" theory.
We intend to use the propagators derived in \cite{ra.} to evaluate the 
contribution of the 1-loop Feynman diagrams to the mean value 
$\overline{Q^{\a\b}}$, 
to the bare propagators $G^{\a\b\c\d}$, and to the bare cubic vertices.  

We consider the 1-loop Feynman diagrams as a 
g-series expansion, by using the correspondent propagators. 
In analogy with the static, we can guess that  
the $g$-dependent part of the 1-loop corrections is not 
singular at the critical point, as soon as $g\neq 0$. 
The physical quantities have  to be not affected by the value of 
$g$, provided that it is not zero. The behaviour at $g=0$ is rather different
and the $\epsilon$-expansion starts from $D=8$, \cite{lu.}. 

Concerning the 1-point function, let us consider the response function, that
as a consequence of eq. (\ref{q}) is: 
 \be
G(\omega)= 2\la Q_{k=0}^{21}(\omega)\ra
\ee

\begin{figure}
\leavevmode
\epsfxsize=0.75\textwidth
\epsfbox{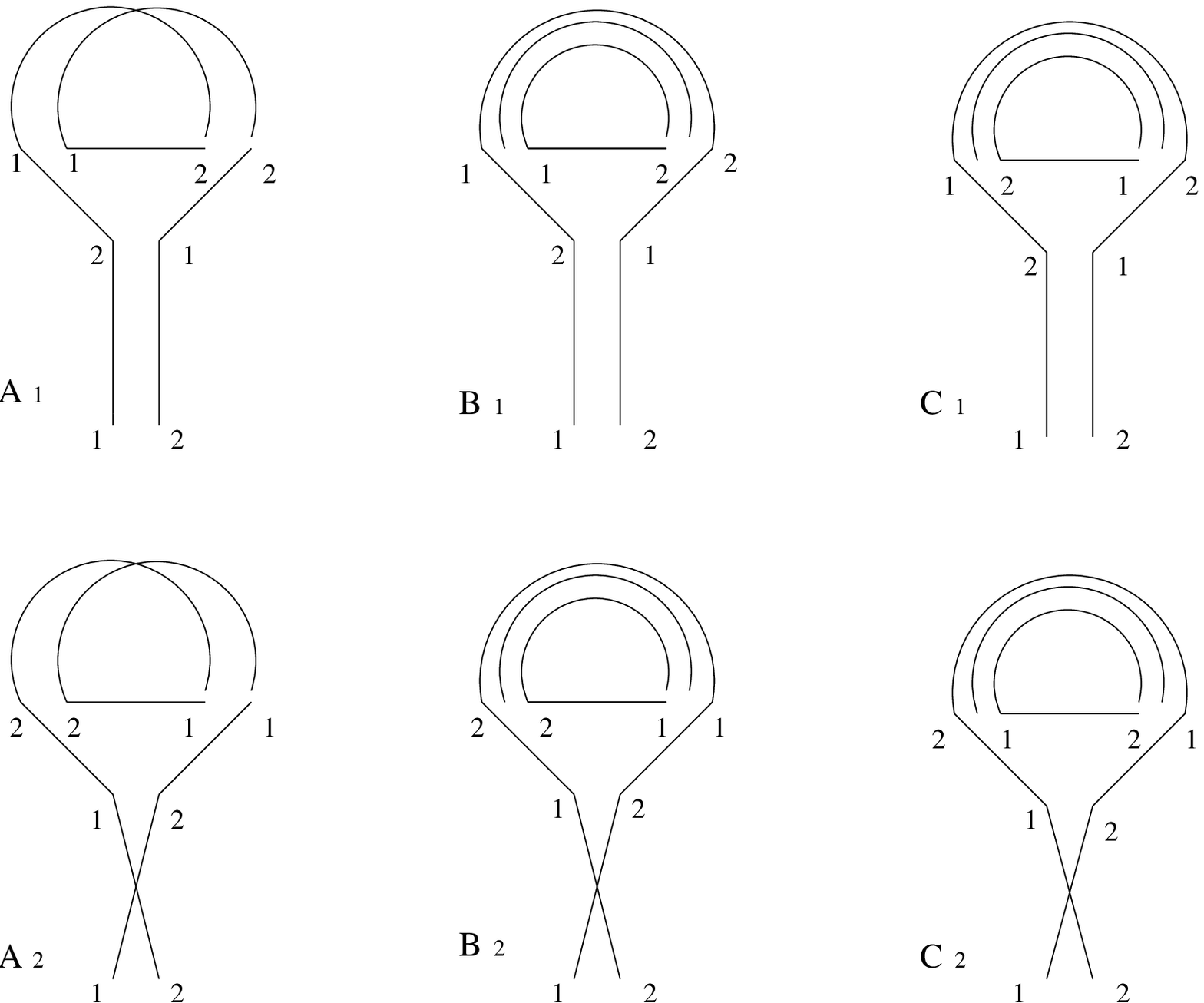}
\caption[]{}
\label{}
\end{figure}

The diagrams that occur in 1-loop correction to the MF value of the $x$
exponent are shown in Fig.[1]:
two continuous lines represent a bare factorized in time  propagator 
$$\tilde{G}^{\a\b\c\d}(k;w_1, w_2) \left[\d (\w_1+\w_3)\d (\w_2+\w_4)
+\d (\w_1+\w_4)\d (\w_2+\w_3)\right],$$ 
three lines the bare connected in time propagator 
$$\tilde{\tilde{G}}^{\a\b\c\d}(k;w_1,w_2,\w_3,\w_4)
\d (\w_1+\w_2+\w_3+\w_4)$$
and, finally, the triangle in the center of the diagrams is a factorized in 
time cubic vertex.  
\begin{figure}
\leavevmode
\epsfxsize=0.75\textwidth
\epsfbox{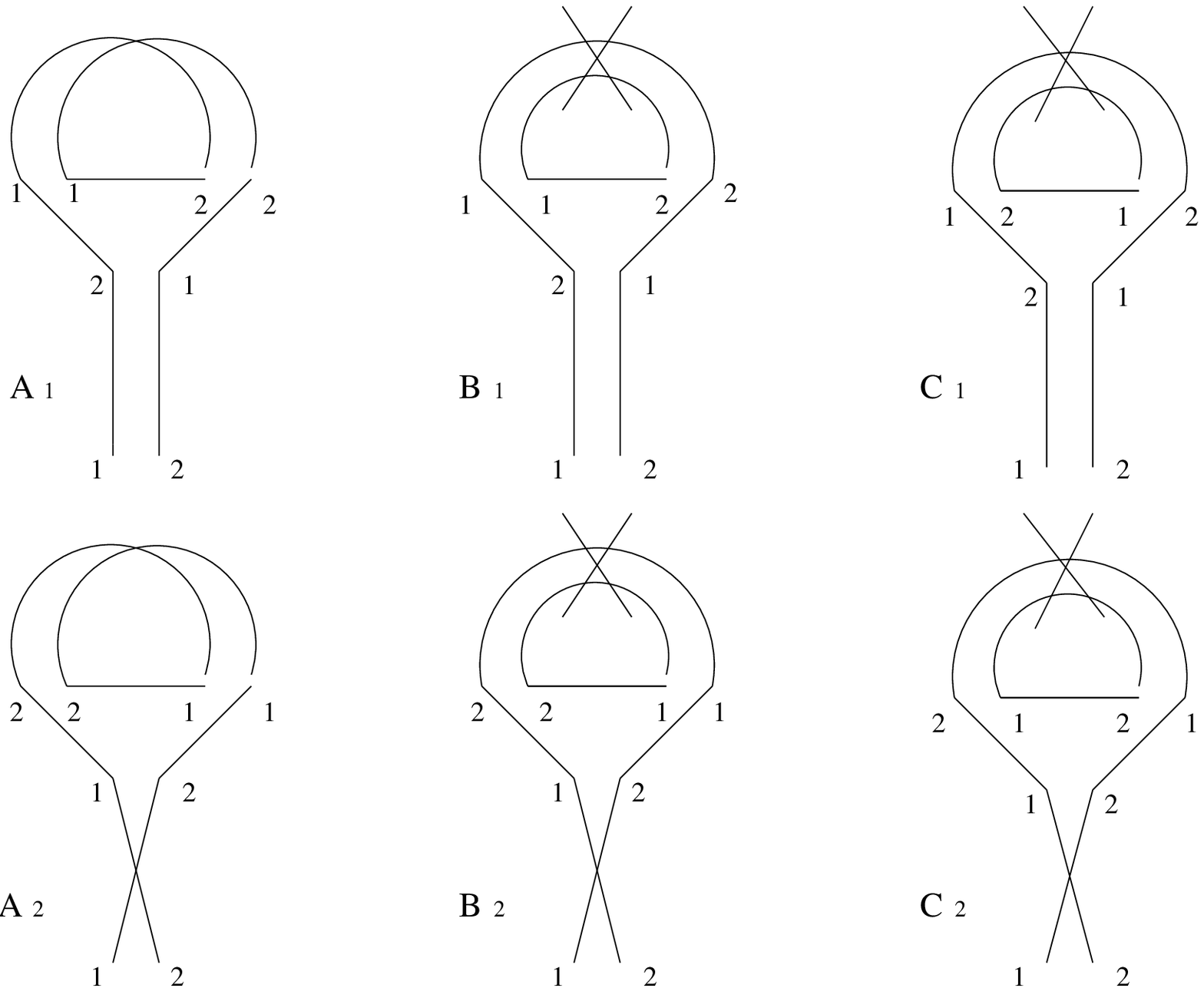}
\caption[]{}
\label{}
\end{figure}
We succeed in evaluating the singular behaviour of the diagrams $A_1$ and 
$A_2$, while we have to use a trick to take into account the contribution 
from the others. 
By using the series expansion in $g$ of the propagators in the loop, 
see \cite{ra.},   
we obtain a correspondent $g$ series of the two 1P.I. diagrams, but
the zero term is missing. 
Let us  add and subtract the term we need. 
As in the static case (where we deal with just the momentum 
variable $k$, and we easily manage to resum the $g$-series),   
we can suppose that the resumation of the series give a non singular 
behaviour at the critical 
point (the presence of $g$ remove the pole at zero momentum for $T=T_c$). 
We left with the diagrams $A_1$, $A_2$ and with the two zero-order in $g$ 
diagrams that we need for the series resumation (with negative sign), 
(see Fig.[2]):
\be
\left(\frac{u}{2}\right)\; 2
\left[\int \frac{d^{d}k}{(2\pi)^d}\frac{1}{ck^2+2\sqrt{i\w}}-2
\int \frac{d^{d}k}{(2\pi)^d}\frac{1}{ck^2+\sqrt{i\w}}\right]
\ee
The propagators involved in these "new" diagrams are represented, in Fig.[2], 
with two crossed lines.
By evaluating the previous integrals for $d=6$, we find that, at the first 
order in the loop expansion, the response function is
\bea
G(\w)&=&1-(-i\w)^x=\overline{G}(\w)+2\;u\;\la\delta Q^{21}(\w)\ra\nonumber\\
&=&1-\sqrt{-i\w}-\frac{k_6} {(2\pi)^6}u^2
 \left[I(\w=0)+\frac{1}{4}\sqrt{-i\w}\:\ln(\w)\right]
\label{g3}
\eea
As for the statics $u$ has been introduced as expansion parameter in the 
cubic vertices, to apply the renormalization group method for critical 
phenomena. 
The factor $u^2$ in the equation (\ref{g3}) is due to the fact that 
the saddle point response function (\ref{g2}) is of order $(\frac{1}{u})$.

In the same way, we derive 
the flux equation of the 1P.I. vertex functions which allow us to
determine the fixed point below $6$ dimension.
As for the conventional Langevin dynamical theories, the IR stable fixed point 
below $6$ dimension is the same that the static one (i.e., we find the same
relevant diagrams):
\be
\label{fp}
(u^*)^2=\frac{(2\pi)^6}{k_6}\frac{\eps}{2}.
\ee
The series expansion in $\eps$ for the static exponents is evaluated up the
$3$th order in Ref. \cite{c.l.} and Ref. \cite{gr.}.

Let us determine the first order correction for the dynamical ones.
Substituting the fixed point value (\ref{fp}) into (\ref{g3}) we obtain for
$x$:
\be
x=\frac{1}{2}\left(1-\frac{\eps}{4}\right).
\ee
In the same way, we find that the relevant 1-loop 1P.I. diagrams for 
the Self-Energy $\Sigma(\w_1, \w_2)$, shown in Fig [3], give the following 
contribution to the spin glass susceptibility (\ref{su2}):
\begin{figure}
\leavevmode
\epsfxsize=0.75\textwidth
\epsfbox{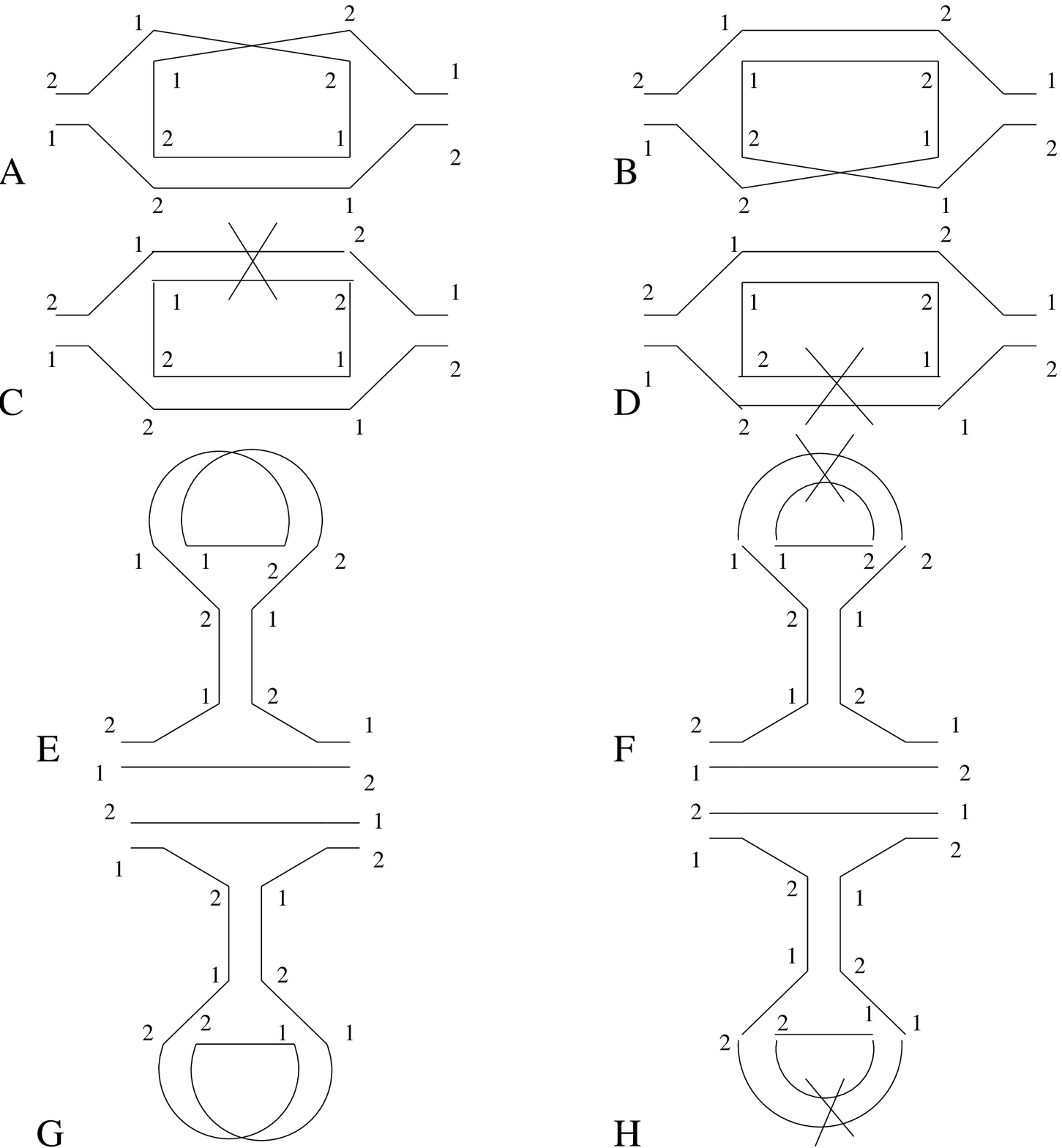}
\caption[]{}
\label{}
\end{figure}
\bea
&&\displaystyle{
A:\frac{u^2}{2}\int\frac{d^{d}k}{(2\pi)^d}\left(\frac{1}{c(p-k)^2
+2\sqrt{i\w_1}}\frac{1}{ck^2+\sqrt{i\w_1}+\sqrt{-i\w_2}}\right)}\\
&&\displaystyle{
B:\frac{u^2}{2}\int\frac{d^{d}k}{(2\pi)^d}\left(\frac{1}{c(p-k)^2
+2\sqrt{-i\w_2}}\frac{1}{ck^2+\sqrt{i\w_1}+\sqrt{-i\w_2}}\right)}\\
&&\displaystyle{
C:-2\frac{u^2}{2}\int\frac{d^{d}k}{(2\pi)^d}\left(\frac{1}{c(p-k)^2
+\sqrt{i\w_1}}\frac{1}{ck^2+\sqrt{-i\w_2}}\right)}\\
&&\displaystyle{
D:-2\frac{u^2}{2}\int\frac{d^{d}k}{(2\pi)^d}\left(\frac{1}{c(p-k)^2
+\sqrt{-i\w_2}}\frac{1}{ck^2+\sqrt{i\w_1}}\right)}\\
&&\displaystyle{
E:\frac{u^2}{2}\; 2\;\frac{1}{cp^2+\; 2\sqrt{i\w_1}}
\int \frac{d^{d}k}{(2\pi)^d}\frac{1}{ck^2+2\sqrt{i\w_1}}}\\
&&\displaystyle{
F:-2 \frac{u^2}{2}\; 2\;\frac{1}{c p^2+\; 2\sqrt{i\w_1}}
\int\frac{d^{d}k}{(2\pi)^d}\frac{1}{ck^2+\sqrt{i\w_1}}}\\
&&\displaystyle{
G:\frac{u^2}{2}\; 2\;\frac{1}{cp^2+\; 2\sqrt{-i\w_2}}
\int \frac{d^{d}k}{(2\pi)^d}\frac{1}{ck^2+2\sqrt{-i\w_2}}}\\
&&\displaystyle{
H:-2 \frac{u^2}{2}\; 2\;\frac{1}{cp^2+\; 2\sqrt{-i\w_2}}            
\int\frac{d^{d}k}{(2\pi)^d}\frac{1}{ck^2+\sqrt{-i\w_2}}}
\eea
At the first order in  $\eps$, the sum of the contributions 
to $z$ exponent of the diagrams A, B, C, D is vanishing. 
On the other hand, from the sum of the diagrams E, F, G and H 
we obtain the following contribution 
to the 1P.I. 2-point function:  
\bea
&&\left.\chi^{-1}_{SG}(p;-\w_1,\w_2)\right|_{p=0}=
\overline{\chi^{-1}_{SG}}(-\w_1, \w_2)+\Sigma(\w_1,\w_2)=\nonumber\\
&&\sqrt{i\w_1}+\sqrt{-i\w_2}+
             \frac{u^2}{4}\frac{k_6}{(2\pi)^6}
\left(\sqrt{i\w_1}\;\ln(\w_1)+\sqrt{-i\w_2}\;\ln(\w_2)\right) 
\eea  

At the fixed point $u^*$ given from (\ref{fp}), we obtain the following 
correction to $z$ exponent:
\be
z=\frac{(2-\eta)}{\frac{1}{2}\left(1+\frac{\eps}{4}\right)}= 
4\left(1-\frac{\eps}{12}\right)
\ee 

The scaling relation between the exponents $x$ and $z$, that we recall to be
\be
x=\frac{d-2+\eta}{2 z},
\ee
is verified, at the first order in $\eps$. 


Numerical simulations for the exponents $x$ and $z$ can be found in 
the literature, the values are $z\approx 7$ and $x\approx .06$ in 
dimensions $D=3$, \cite{og.}, and $z\approx 5$ and $x\approx .15$ 
in dimensions $D=4$, \cite{ru.}. 
Our prediction states that both the values of $x$ and $z$ are decreasing 
when the dimension decreases. This is true for $x$ but not for $z$.
The apparent discrepancy that we have with the behaviour of $z$ 
should not worry us. In fact, also in the static case, the critical 
exponents for spin glasses have a badly convergent $\eps$-expansion and the 
prediction of this expansion can be hardly applied in 3 or 4 dimensions.

A numerical study of what happens in 5 dimensions is necessary. 
Moreover it should be noticed that usual arguments imply 
that our computation predicts, without ambiguities, that the 
logarithmic corrections in $6$ dimensions are such to decrease the 
effective value of the exponents.

\newpage

\begin{table}
\begin{tabular}{c}

CAPTIONS FOR ILLUSTRATIONS

\vspace{1cm}

\end{tabular}

\begin{tabular}{ll}
Fig.1 & 1-loop diagrams that contribute to the $x$ exponent.

\vspace{.5cm} \\

Fig.2 & g-series resumation of the 1-loop diagram contribution to
        the $x$ exponent.

\vspace{.5cm} \\

Fig.3 & g-series resumation of the 1-loop diagram contribution to the
       Self-Energy.\\

\vspace{10cm}

\end{tabular}
\end{table}
\newpage 
\end{document}